\documentclass[letterpaper,twocolumn,10pt]{article}

\usepackage{epsfig,endnotes, multirow}
\usepackage{algorithmic, algorithm}
\usepackage{graphicx}
\usepackage{blindtext}
\usepackage{pdfpages}
\usepackage{color}
\usepackage{bbm}
\usepackage{subcaption}
\usepackage{fancyvrb}
\usepackage{cleveref}
\usepackage{cite}
\usepackage{enumitem}
\usepackage{pgf-pie}
\usepackage{amsthm}
\usepackage[toc,page]{appendix}
\usepackage{xcolor,colortbl}
\usepackage{soul}

\crefformat{footnote}{#2\footnotemark[#1]#3}

\definecolor{mygreen}{rgb}{0,0.5,0}
\definecolor{myred}{rgb}{0.5,0,0}
\definecolor{myblue}{rgb}{0,0,1}
\definecolor{myorange}{rgb}{1,0.65,0}

\newcommand\ignore[1]{}

\usepackage{epsfig,endnotes}

\usepackage{rotating}

\newenvironment{Mydescription}
               {\list{}{\labelwidth 0pt \leftmargin=0pt \itemsep=2pt
                        }}
               {\endlist}

\begin{document}
\date{}


\title{ Eradicating Attacks on the Internal Network with\\
Internal Network Policy }


\author{
{\rm Yehuda Afek  }\\
Tel-Aviv University
\and
{\rm Anat Bremler-Barr}\\
IDC
\and
{\rm Alon Noy}\\
Tel-Aviv University
}






\maketitle

\thispagestyle{plain}
ֿ

\begin{abstract}
In this paper we present three attacks on private internal networks behind a NAT and a corresponding new protection mechanism, Internal Network Policy, to mitigate a wide range of attacks that penetrate internal networks behind a NAT. In the attack scenario, a victim is tricked to visit the attacker's website, which contains a malicious script that lets the attacker access the victim's internal network in different ways, including opening a port in the NAT or sending a sophisticated request to local devices. The first attack utilizes DNS Rebinding in a particular way, while the other two demonstrate different methods of attacking the network, based on application security vulnerabilities.
Following the attacks, we provide a new browser security policy, Internal Network Policy (INP), which protects against these types of vulnerabilities and attacks. This policy is implemented in the browser just like Same Origin Policy (SOP) and prevents malicious access to internal resources by external entities.
\end{abstract}

\section{Introduction}
A large class of malicious exploits on the internet trick users to transmit an undesired request to a website that trusts them, CSRF (Cross Site Request Forgery) {\cite{csrf}} is an example sub class of these attacks.
A key mechanism to limit the capabilities of these attacks is the SOP (Same Origin Policy) {\cite{sop}}, which restricts the type of scripts that one site can execute on the browser, in the context of a different site. Recently, attacks with the same general principle have been presented, causing the victim's browser to access IoT devices and other resources on the victim's private network, behind a NAT \cite{acar2018web,dns_rebinding_private_networks}. 

We review the previously known attack by Acar et al. {\cite{acar2018web}} on an internal network that uses DNS Rebinding {\cite{kokkinopoulos2009dns}} to circumvent the SOP through sophisticated mapping of DNS entries to IoT devices in the internal network. This attack lets the attacker gain access to IoT devices only by tricking the victim to surf to a malicious website.
Following this attack we present three additional attacks, each circumventing SOP in a different way. The first one is similar in its flow to the Acar at el. attack but, unlike it, instead of gaining access to IoT devices, we utilize DNS Rebinding and UPnP (Universal Plug \& Play protocol) {\cite{rfc6970}} to directly attack the home router and gain access to the internal network, thus demonstrating that home routers, which are considered also to be a kind of IoT device, are the weakest point of the internal network.
75\% of our tested home routers, which are supplied by ISPs in our region, are vulnerable to this attack. Next we demonstrate two more attacks on devices behind a NAT which do not use DNS Rebinding. The first targets IoT devices and achieves the same results as the attack of Acar et al. {\cite{acar2018web}}, while the second bypasses previously known application security mitigations {\cite{kombade2012csrf}} to attack home routers.

To circumvent these and similar attacks on devices behind a NAT, we present Internal Network Policy (INP), a new natural complementary extension of the browser security policy, SOP. INP prevents all attack types presented, and enhances the protection and security of private networks. INP can also be used for private networks without a NAT, though it does not provide the same security isolation that a NAT provides. One of the major advantages of implementing defensive solutions in web browsers is the frequency of browser updates {\cite{browsers_update}}, in contrast to network devices such as home routers or IoT devices.

INP stops attacks on devices behind a NAT that trick the victim to surf to a malicious website. There are other attack vectors for penetrating an internal network, such as gaining physical access to the network or exploiting a vulnerability in software which communicates directly with the internet (such as an online chat or e-mail). These other attack vectors are more complicated and less common than attacks that use the web browser's functionality \cite{mozilla_rce_stats, chromevuln2019}. Note that these other attacks are supposedly stopped by IPS and firewall security devices.

\textbf{Paper organization.} 
After reviewing related work (Section \ref{sec:related}), we review the attack by Acar et al. and present three new attacks (Section \ref{sec:attacks}). Each attack highlights a different aspect of the root problem, all of which are solved by INP, the new security policy suggested here. We propose the INP defensive mechanism (Section \ref{sec:inp}), explain how it works and why it protects from attacking behind a NAT, including a proof of concept implementation. In the end we show the experiments (Section \ref{sec:experiments}) that we have done to perform the attacks, check which devices are vulnerable and test the effectiveness of INP. We make parts of our work available online, as described in Section \ref{sec:availability}.
 \section{Related Work}
 \label{sec:related}

\subsection{Attacks on IoT behind NAT}

In their paper, "Web-based Attacks to Discover and Control Local IoT Devices" \cite{acar2018web}, Acar et al. demonstrated an attack on IoT devices in home networks behind a NAT. The attack is based on DNS Rebinding. Other attacks that use DNS Rebinding to attack private networks are presented in Dorsey's article \cite{dns_rebinding_private_networks}. \ignore{There are common guidelines between these attacks to the one shown by Acar et al. \cite{acar2018web}.}

Table \ref{table:related_comparison} summarizes the differences between the attack of Acar et al. and the three new attacks that we present here. Two of the new attacks that we present in Section \ref{sec:attacks} gain access to private networks without using DNS Rebinding. 

\subsection{Preventing attacks on the internal network}

Many defensive approaches have been created to deal with attacking private networks behind a NAT. Some of them tried to solve DNS Rebinding \cite{jackson2009protecting,pandiaraja2015applying} by creating specific rules in the home routers \cite{dnswall}. Most of the given approaches were difficult to implement and integrate with network devices. Moreover, many other application security mitigation concepts were invented, but vulnerabilities are still found and exploited, especially in IoT devices \cite{iotvulns}.

Johns et al. present extended Same Origin Policy, "eSOP" \cite{johns2013eradicating}. Johns et al. analyze the weaknesses of SOP and indicate DNS Rebinding as the main problem which should be solved. eSOP was invented to combat DNS Rebinding by using new headers which should be supplied by web servers in HTTP responses. In our paper, we analyze the problem of attacking internal network behind a NAT differently, and expose a different new root problem. A solution that is based only on DNS Rebinding cannot prevent our second and third attacks, nor does eSOP stop our first attack. The INP presented here copes with both DNS Rebinding and other attacks, including the new one presented in this paper. INP is secure by default, which means that if a vendor's device is not compatible with it, external access to the device is prevented. Table \ref{table:related_comparison} provides in addition a comparison of eSOP and SOP relative to these attacks.

Note that our first attack (Subsection \ref{sec:attack1}) bypasses eSOP, even though the attack uses DNS Rebinding. Although eSOP prevents an attack script from reading HTTP responses from other websites, our first attack needs nothing but sending a UPnP command in an HTTP request to the home router, the response is ignored. 

Another IoT defensive approach is the concept of MUD (Manufacturer Usage Description) {\cite{hamza2018clear_as_mud}}, which lets IoT vendors provide a type of white-list specifying their device legitimate network behavior which can then be monitored and enforced by appropriate security devices. However, MUD relies on various IoT vendors to provide the MUD file. Moreover, the attacks shown in this paper, are not prevented by MUD, as they exploit the local HTTP servers of the IoT devices, which should legitimately be used in the LAN.

\begin{table*}[h!]
\centering
\begin{center}
\resizebox{\textwidth}{!}{\begin{tabular}{|l|l|l|c|c|}
\hline
 & \textbf{Building Blocks} & \textbf{Outcome} & \textbf{eSOP mitigate} & \textbf{INP mitigate}  \\
 \hline
 \textbf{Acar et al.} & DNS rebinding & Control IoTs & \cellcolor{green!25}Yes & \cellcolor{green!25}Yes \\
 \textbf{Attack I} & DNS rebinding, Preflight action & UPnP on home router & \cellcolor{red!25}No & \cellcolor{green!25}Yes \\
 \textbf{Attack II} & CSRF & Control IoTs & \cellcolor{red!25}No & \cellcolor{green!25}Yes \\
 \textbf{Attack III} & Network proximity, XSS  & Control home router & \cellcolor{red!25}No & \cellcolor{green!25}Yes \\ 
 \hline
 \end{tabular}}
\end{center}
\caption{Attacks comparison and mitigation techniques comparison. }
\label{table:related_comparison}
\end{table*}

While implementing INP on Chromium, we noticed that in 2017 Mike West offered modifications to Chromium {\cite{cors_rfc1918} that head in the same direction as INP, preventing external resources from accessing internal network devices using the browser as a stepping stone. However, we believe the INP presented here addresses a wider range of issues not considered by the modifications of \cite{cors_rfc1918}, including (i) dynamic and automatic classification of internal IP address space by the browser, dealing with cases where the private IP addresses are not only according to RFC 1918 \cite{rfc1918}. This is necessary in bigger organizations, such as large companies or universities, which do not always use RFC 1918 addresses for their internal networks. Furthermore, with the growing support for IPv6 many use cases for RFC 1918 addresses cease to exist \cite{johns2013eradicating}, (ii) INP as suggested here provides configuration API rules that network administrators can manually add internal IP addresses on top of the default that should include the RFC 1918 ranges, (iii) our INP preflight request provides details about the IP address of the request initiator, thus enabling vendors of network devices (IoT, home routers, etc.) to allow cross site references if they come from specific external IP addresses such as the IoT vendor site, and (iv) the modifications proposed by West do not always prevent attacks whose HTTP requests' destination hosts are given as domain names, rather than IP addresses, when the domain name has not yet been resolved. This still enables attacks such as those relying on DNS Rebinding. Our INP implementation (see Subsection \ref{sec:implementation}) handles these cases and protects internal resources from being accessed through domain names.
 \section{Attacks behind a NAT}
 \label{sec:attacks}
Four attacks that execute a script (originated in the attacker's website) on the web browser in order to access a device in the internal network are reviewed here. First we provide a basic background for SOP (Same Origin Policy). Then, we review the attack of Acar et al. \cite{acar2018web}, and go on to present three new attacks.
\subsection{ Same Origin Policy }
\label{sec:sop}
SOP (Same Origin Policy) \cite{sop} is a standard security policy used by most browsers in order to prevent a malicious script on one page from obtaining access to sensitive data on another web page through that page's Document Object Model (DOM). An origin is defined as a combination of URI scheme, host name and port number. Under the policy, a web browser permits scripts contained in one web page to access data in a second web page only if both web pages have the same origin. Static resources such as images or frames may be embedded cross-origin. 

Two major rules of SOP are the following restrictions:

\begin{itemize}
    \item \textbf{Blocking the response.} If a script sends an HTTP request to a site of a different origin (i.e., cross origin), SOP prevents the script from reading the response, unless the other site allows it with corresponding fields set in the HTTP response headers \cite{cors}.
    \item \textbf{Deny complex requests.} An HTTP request is considered complex if it contains customized HTTP headers, or non-trivial textual content type (otherwise the request is considered simple). If a script sends a cross origin complex request, SOP blocks that request by default. Instead, SOP instructs the browser to send a preceding \textbf{preflight request} \cite{cors_preflight}, asking the other site whether it allows the original request to be sent to it. If the other site allows that with corresponding fields set in the headers in the response to the preflight request, then the browser sends the original request.
\end{itemize}
\subsection{ Background: Acar et al. attack on IoTs \cite{acar2018web} }
\label{sec:attack_acar}

The attacker's goal in the attack presented by Acar et al. is to access local devices in the victim's network in order to extract information from or send commands to the devices. The script which instructs the browser to make that access is placed in the attacker's evil.com website. First, the attacker tricks the premise user to visit its website using the script with the bait. A direct request to the IoT device points to a different origin (the IP address of the device is different than the address of evil.com), so the script would not be able to access the response. In order to perform the request on the local IoT device and overcome the SOP, DNS Rebinding is used, switching the IP of evil.com to the IP of the desired internal device.

The following is a step by step detail of the attack (see Figure \ref{fig:attack_acar_flow}):
\begin{enumerate}
    \item \textbf{The victim surfs to the attacker's website evil.com.}
    \begin{enumerate}
        \item The browser tries to resolve the IP address of evil.com. The attacker's controlled response points to its external web server, with low TTL value (e.g. one second).
        \item The victim sends an HTTP GET request to evil.com, which responds with the attacker's malicious script.
    \end{enumerate}
    \item \textbf{The malicious script repeatedly sends GET requests to http://evil.com/evil-test.}
    Until the evil.com entry in the victim DNS cache is evicted (due to expired TTL) the request keeps going to the attacker's website, which responds with HTTP 200 OK (indicating that the request has succeeded) responses. 
    \item \textbf{Attacker changes evil.com DNS record.}
    When the entry's TTL expires, DNS Rebinding by the attacker replaces the evil.com entry to point to the internal IP address of the desired device.
    \item \textbf{DNS Rebinding is completed.}
    The victim tries again to resolve evil.com, which is now resolved to the IP address of the desired internal device. 
    \item \textbf{Local device is accessed by the attacker.}
    The attack script can now send HTTP requests directly to the HTTP server on the device and read responses, allowing the attacker to extract information from or send commands to the device.
\end{enumerate}

\begin{figure}[h]
   \centering
   \begin{tabular}{@{}c@{\hspace{.5cm}}c@{}}
       \includegraphics[page=12,width=.45\textwidth]{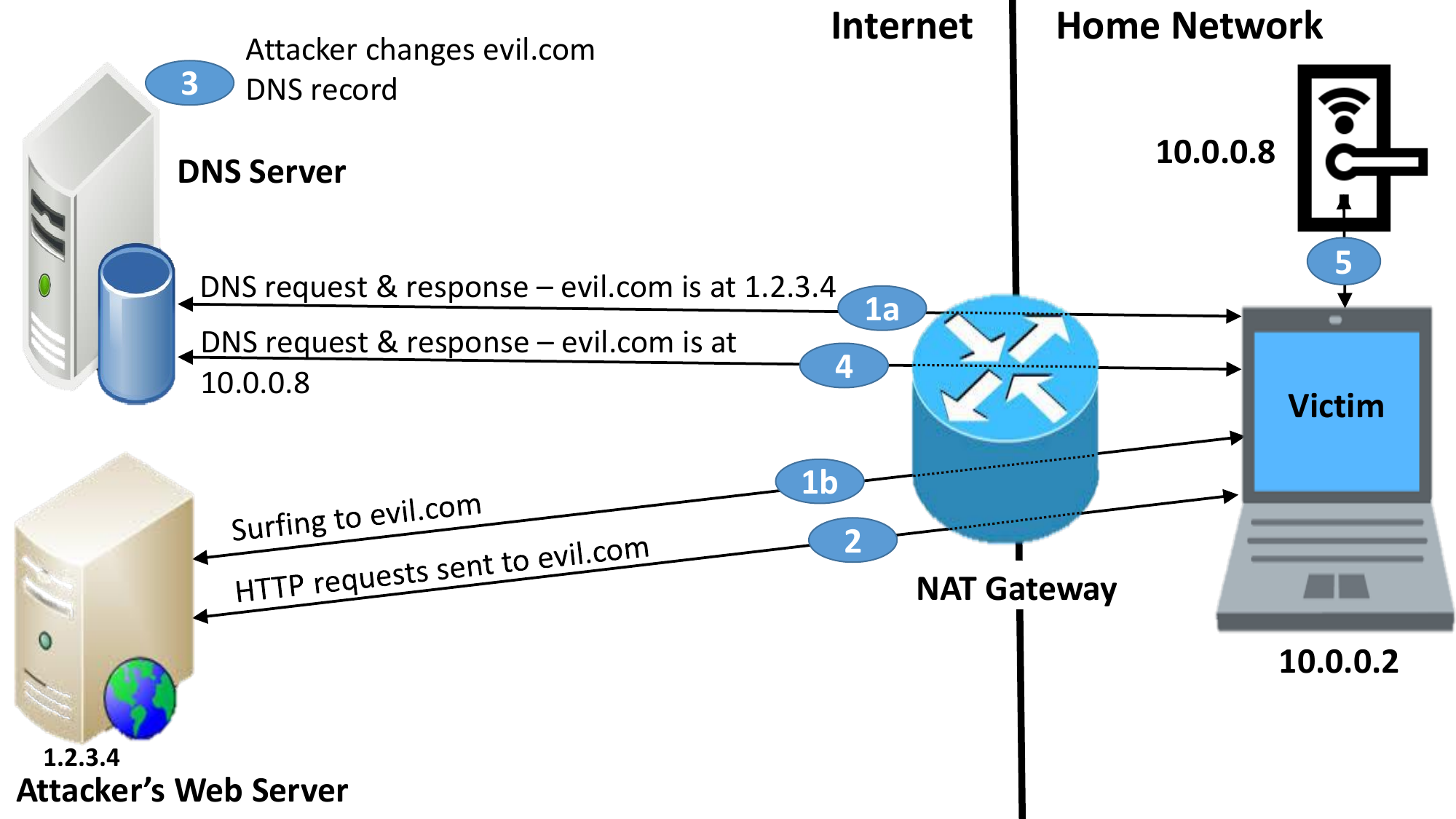}
   \end{tabular}
 \caption{Flow of Acar et al. attack}
 \label{fig:attack_acar_flow}
\end{figure}

In their paper \cite{acar2018web}, Acar et al. use an internal scan before the attack in order to detect specific IoT devices. Later they use the results of the scan to execute the presented attack.
\subsection{ Attack I: Attacking home routers with DNS Rebinding \& UPnP }
\label{sec:attack1}

In this attack, the goal is to open a port in the NAT on the gateway (the victim's home router). It lets the attacker access practically any device in the network without any prior scan. The instruction to open the port is placed in a script that is located in the attacker's evil.com website. Most of the attack steps are similar to the previous ones presented by Acar et al. Nevertheless, this attack focuses directly on the home router, the single critical point of failure in the network. The attack is effective on 75\% of the routers that are supplied by ISPs in our region, which we have verified (vulnerable devices and experiments are detailed in Section \ref{sec:experiments}).

Together with DNS Rebinding, this attack takes advantage of the UPnP (Universal Plug \& Play) \cite{rfc6970} protocol, used by most network devices, including home routers. A router's UPnP server supports HTTP requests which contain commands, such as adding a port forwarding rule or changing the DNS server.

The full attack flow can be viewed in Figure \ref{fig:attack1flow}.

\begin{figure}[h]
   \centering
   \begin{tabular}{@{}c@{\hspace{.5cm}}c@{}}
       \includegraphics[page=6,width=.45\textwidth]{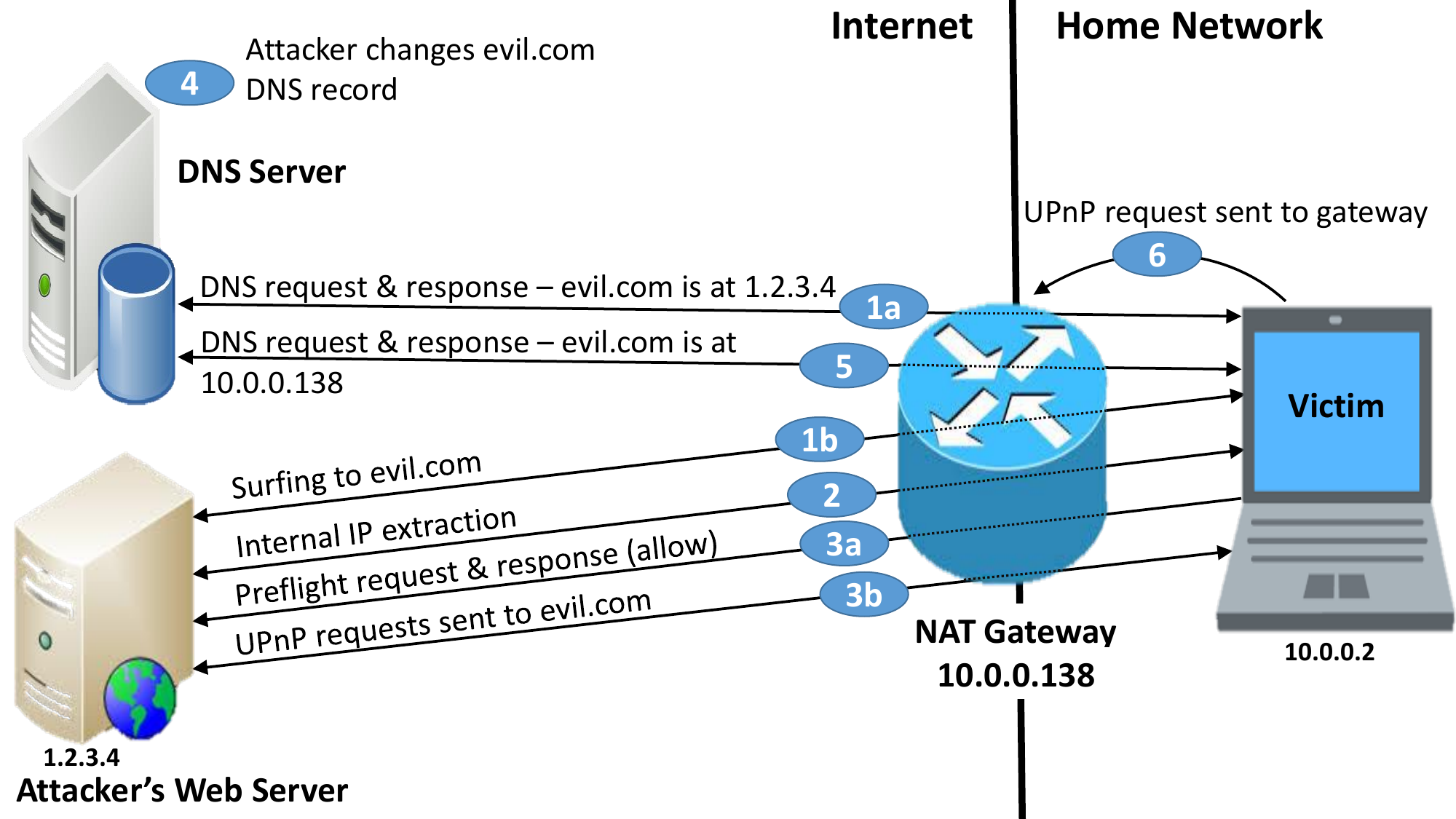}
   \end{tabular}
 \caption{Attack I Flow}
 \label{fig:attack1flow}
\end{figure}

\begin{enumerate}

\item \textbf{The victim surfs to the attacker's controlled website, evil.com.}
\begin{enumerate}
    \item The same as in the Acar et al. attack above.
    \item The same as in the Acar et al. attack above.
\end{enumerate}

\item \textbf{Extracting the victim's gateway internal IP address.}
The WebRTC \cite{webrtc} extension allows JavaScript to query the local client IP address. Acquiring the victim's internal IP address lets the attacker predict with a good probability of success the internal IP address of the gateway. Most of the networks share the same standards.

\item \label{item:sendUPNPP} \textbf{The malicious script repeatedly sends UPnP AddPortMapping requests to evil.com.}
\begin{enumerate}
    \item Unlike the requests sent in the Acar et al. attack, UPnP HTTP requests are used here:
     \begin{itemize}
         \item Destination port may be different than 80 - which means that the UPnP request is considered cross origin.
         \item The content type is XML - which means the UPnP request is considered complex (as explained in Subsection \ref{sec:sop}). 
     \end{itemize}
     Therefore, preflight \cite{cors_preflight} requests are triggered and sent to evil.com, which responds with corresponding "allow" headers.
     Because the attacker lacks any prior knowledge about the victim's router, the attack script sends the UPnP command to evil.com to all the known UPnP ports in parallel (tested UPnP ports and URLs are demonstrated in Appendix \ref{sec:appendix1}).
    \item The victim's browser now repeatedly sends UPnP AddPortMapping requests, as instructed in the script, to the web server at evil.com, which ignores the content and replies with HTTP 200 OK.
\end{enumerate}

\item \textbf{Attacker changes the evil.com DNS record.}
It is the same as in the Acar et al. attack above, but it replaces the evil.com entry to point to the internal IP address of the router.

\item \textbf{DNS Rebinding is completed.}
It is the same as in the Acar et al. attack above. The victim resolves evil.com again, which is now resolved to the internal IP address of the gateway. 

\item \textbf{UPnP AddPortMapping request is sent to the gateway.}
The attacker's script is still executed in the victim's browser, thus the victim sends the UPnP requests to evil.com, which is now the victim's home router. The attack goal is now achieved, a port forwarding rule is added, and the attacker can directly access the home network.
\end{enumerate}

Except for the third step, most of the flow is very similar to the flow of Acar et al. In comparison to the attack by Acar et al., an initial scan of the internal network is not necessary.
The ability to add port forwarding rules to the home router is actually the significant step in taking over the victim's network, as local devices can be accessed directly. By using a RCE (Remote Code Execution) vulnerability \cite{bluekeep} or simple credentials, which are still common in many home devices \cite{bad_router_passwords}, the attacker can gain control of the entire internal network.

In Section \ref{sec:experiments} we show that 75\% of the examined routers, supplied by the ISPs in our region are vulnerable to our attack.
In the demonstrated attack flow, we target known UPnP servers whose URLs and TCP ports are fixed and static.
In Appendix \ref{sec:appendix1} we show how we are able to attack routers whose UPnP TCP port is randomized and dynamically changes.
Routers whose UPnP URLs contain a long, random identifier, such as UUID, are not vulnerable as the attacker cannot predict the URL to target.
\subsection{ Attack II: Attacking internal devices without DNS Rebinding }
\label{sec:attack2}
The second attack is based on the observation that SOP allows a site from one origin to send simple HTTP requests to a site from a different origin with no preceding preflight request (in contrast to complex cross origin requests). That means that if an IoT device can be controlled only by receiving a simple HTTP request, an attack script can exploit it, without DNS Rebinding. The request may utilize the device API, or in some cases exploit a vulnerability such as BOF (Buffer Overflow) \cite{chen2018iotfuzzer}, in order to execute code on the device. 

Therefore, this attack demonstrates that a holistic solution to attacks behind a NAT, cannot be solved, by just preventing DNS Rebinding.

In Figure \ref{fig:attack2flow}, an example of such an attack is demonstrated.

\begin{figure}[h]
   \centering
   \begin{tabular}{@{}c@{\hspace{.5cm}}c@{}}
       \includegraphics[page=8,width=.45\textwidth]{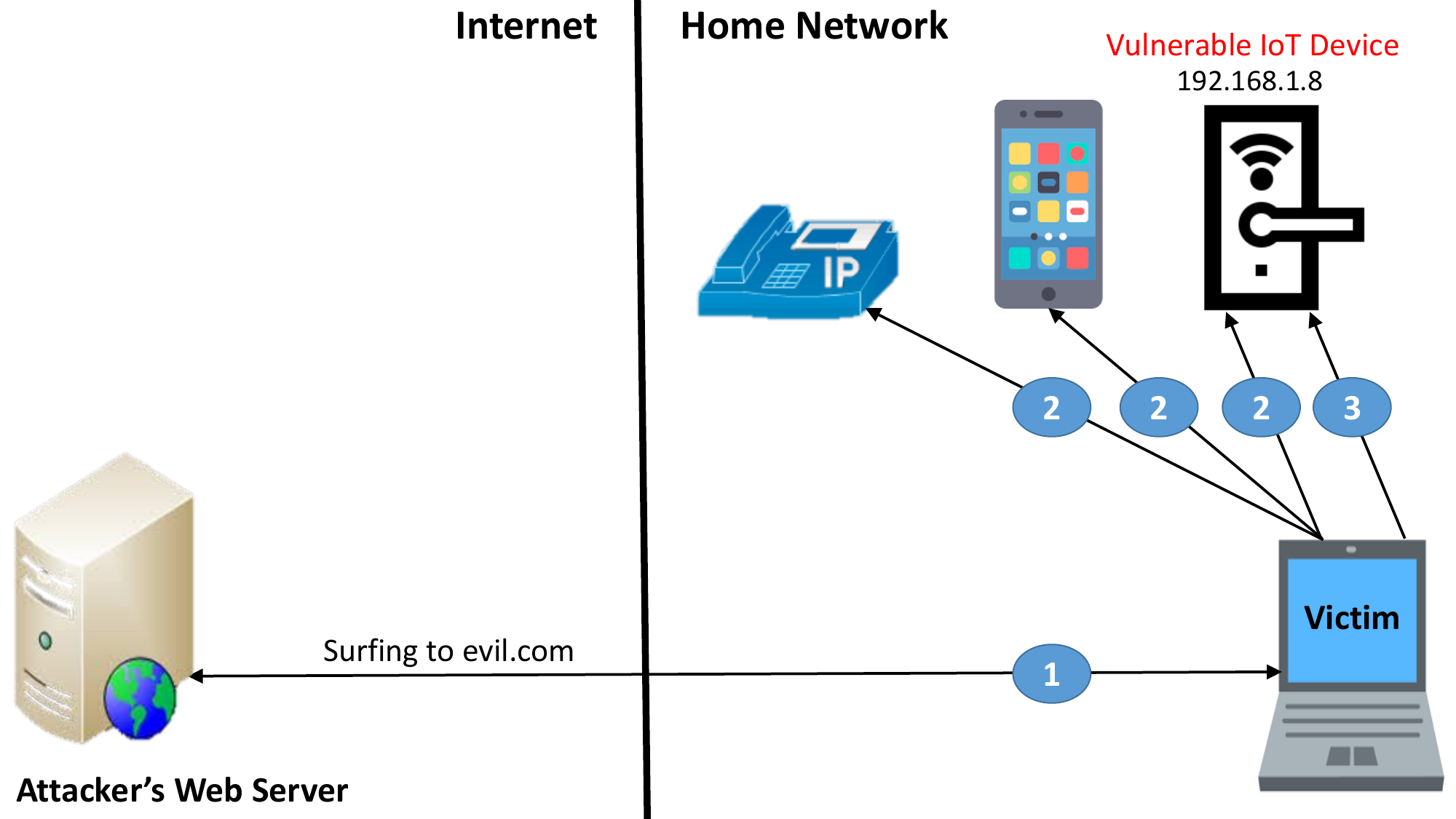}
   \end{tabular}
 \caption{Attack II Flow}
 \label{fig:attack2flow}
\end{figure}

\begin{enumerate}
\item The victim is tricked to surf to the attacker's controlled website, evil.com - the browser retrieves the malicious script.
\item Through the malicious script, the attacker first scans the local network, to discover vulnerable devices (technical details available in Appendix \ref{sec:appendix1}). Let us assume a vulnerable device is found with internal IP address 192.168.1.8.
\item The script sends a simple \cite{cors} HTTP request to the device at 192.168.1.8, and completes the attack.
\end{enumerate}

This attack may let an attacker remotely control IoT such as multimedia devices (smart TVs, AV receivers) by their web management API, or even send crafted HTTP requests to exploit IoT vulnerabilities such as BOF (Buffer Overflow) \cite{belkin_vul}, which would lead to remote code execution on the device. In our experiments, we provide few examples of vulnerable IoT devices (see Section \ref{sec:experiments}).
\subsection{ Attack III: Attacking home routers through static HTML elements } 
\label{sec:attack3}
This attack lets the attacker control the victim's home router by accessing the router's web management interface through the victim's browser. Besides the familiar requirement of tricking the victim to surf to the attacker's malicious evil.com web site, the attack also requires geographical proximity of an attacking machine to the victim. This prerequisite is explained in the sequel.

This attack is capable of making the following penetrations:
\begin{itemize}
    \item Enabling the victim's browser to access the home router by using \textbf{static HTML elements} instead of dynamic scripts. Therefore, SOP and other protection mechanisms such as the NoScript Firefox extension \cite{noscript} do not stop it since they are applied only to dynamic script elements. That proves that attacks behind a NAT could be performed in different ways than executing a script on the victim's browser.
    \item It enables the attacker to \textbf{send any type of HTTP request (including complex ones) to the victim's router without DNS Rebinding}. In comparison to the previous attacks, attack I (Subsection \ref{sec:attack1}) sent UPnP requests, which are complex, but DNS Rebinding was required. Attack II (Subsection \ref{sec:attack2}) didn't use DNS Rebinding, however it enabled the attacker to send only simple HTTP requests to the victim's devices in the internal network.
\end{itemize}

At a high level, this attack exploits a vulnerability in the AP (Access Point) List feature of home routers.

Router web interfaces which include this feature expose a web page with an HTML table that displays Wifi networks that are accessible to the router. One of the table's columns often displays the networks' names (known as SSID), which are derived from networks' routers. With sufficient geographical proximity to the victim, an attacker who sets a Wifi hotspot can control one of the entries in the victim's router AP List table. A basic Wifi hotspot (e.g., created by a smartphone), would be accessible to a router from a maximum distance of less than hundred meters. However, depending on the attacker's antenna and transmitting abilities, it can be accessible from hundreds of kilometers \cite{pietrosemoli2008setting}.

If the attacker sets its network name to be an HTML script tag, such as \begin{verbatim}<script src="http://evil.com/evil.js"/>\end{verbatim} and the victim's router AP List web page does not encode or handle networks' names properly, then the attacker's network name would be interpreted as native HTML. 

If the victim surfs to evil.com while the malicious hotspot is accessible to the victim's router, the only thing left for the attacker to complete the attack is to redirect the victim from evil.com to the router's AP List page. Then the attacker's controlled script (located at http://evil.com/evil.js) is executed on the victim's browser. The most significant consequence is that the attacker's script is executed in the context of the victim's router AP List page, which means that it can freely access the router's web interface without any restriction by SOP.

This vulnerability belongs to a large class of vulnerabilities, named XSS (Cross Site Scripting) \cite{xss}. Previous similar XSS vulnerabilities in home routers have been shown \cite{ssid_xss_blackhat}, however they required the victim to surf directly to the router vulnerable page (instead of evil.com), which dramatically decreases the probability of success.

Following is a step by step detail of the attack (see Figure \ref{fig:attack3flow}):

\begin{figure}[h]
   \centering
   \begin{tabular}{@{}c@{\hspace{.5cm}}c@{}}
       \includegraphics[page=9,width=.45\textwidth]{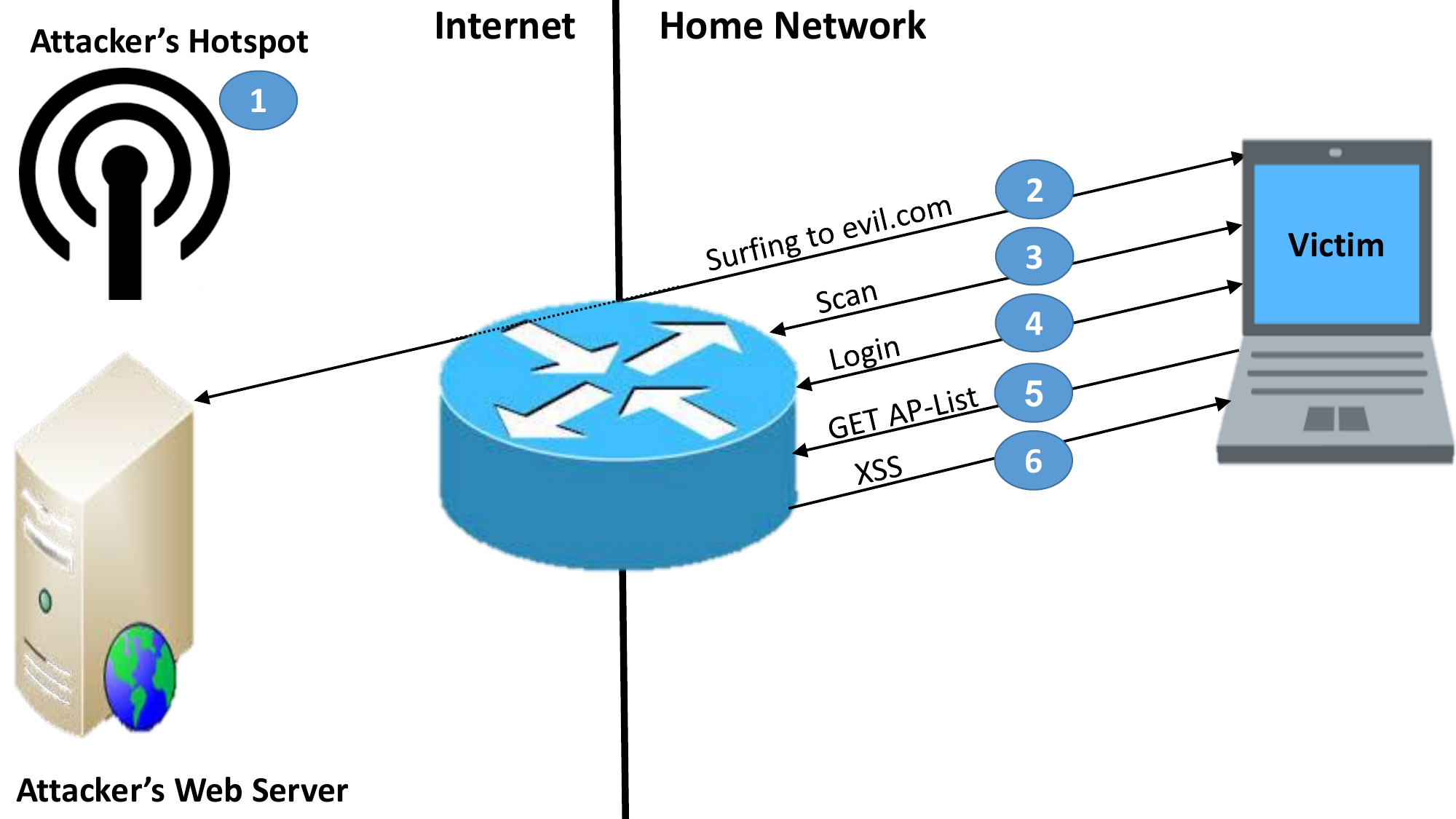}
   \end{tabular}
 \caption{Attack III Flow}
 \label{fig:attack3flow}
\end{figure}

\begin{enumerate}
\item The attacker sets up an access point, with network name: \begin{verbatim}<script src="http://evil.com/evil.js"/>\end{verbatim}
The access point should be accessible by the victim's router.
\item The victim surfs to the attacker's controlled website, evil.com - the browser retrieves the malicious HTML page.
\item The malicious HTML identifies the victim's router, taking advantage of an HTML image element scan technique, as explained in \cite{img_scan} (additional details are available in Appendix \ref{sec:appendix1}).
\item The attacker's HTML makes a login request using an HTML form (see \cite{forms} for details), to the home router. We assume the credentials are default ones (as discussed in \cite{router_passwords}). This makes the victim's session the active session in the router.
\item The attacker's HTML redirects the victim by setting the DOM document location (see \cite{doc_location} for details) to the AP List panel page.
\item The XSS is invoked. The attacker's access point's network name is now parsed as HTML, and a prepared script (at http://evil.com/evil.js) is ready to be executed, in the context of the router itself.
\end{enumerate}

Later in our experiments (Section \ref{sec:experiments}), we show that 41\% of the tested routers supplied by ISPs in our region are vulnerable to this attack.
\section{ INP - Internal Network Policy }
\label{sec:inp}
We suggest Internal Network Policy (INP), a security policy that places strong restrictions on web browser accesses to an internal network device from an external web page, i.e., a web page whose source IP address is external to the network in which the browser is executing. INP is complementary to SOP, and should be used in addition to it. While SOP deals with cross origin access, INP restricts external access to internal IP addresses.

The motivation for INP is the common principle of all the attacks on devices behind a NAT that were discussed in Section \ref{sec:attacks} and other similar attacks. The key principle is the usage of a local web browser as a stepping stone for attacks originating from an external server, on a local device. INP is a security policy that prevents this type of attacks through the browser. 

At a high level consider two disjoint IP address ranges, $I$ (for internal) and $E$ (for external). A web page is called {\em external} if that web page is serviced/downloaded from a server with an IP address in $E$. A resource or web server, is called {\em internal} if its IP address is in $I$.  

A web browser that implements INP permits scripts contained in an external web page to make a request, of any kind, to an internal server only if the internal server pre-approves the request. Pre-approval is implemented with a preflight request-response exchange, similar to the SOP preflight mechanism.
\newtheorem{mydef}{\textbf{Definition}}
\begin{mydef}\label{external_request}
A \textbf{cross-network request} is an HTTP request that is initiated by an external web page to an internal resource.
\end{mydef}

\subsection{INP detailed description}
In our INP implementation the set of internal IP addresses $I$, that is used in order to detect cross-network accesses, is determined in a dynamic way with an option for the system administrator to edit the set.

\ignore{\begin{mydef}\label{internal_ip_address_range-def} 
{\bf Internal IP address range, $I$} the range of IP addresses that are used in a certain premise, for example, in a home, or in an organization, etc. These IP addresses (also called private IP addresses) are used only inside the local network of the premise and may not be used out in the external network, i.e., outside the premise (the same IP addresses may be used by other premises). Any packet with an internal source IP address (as source) that is sent out goes through Network Address Translation which maps it to an external IP address that is route-able in the global Internet.\end{mydef}}

\begin{mydef}\label{internal_ip_address_range-spec-def}
The \textbf{internal IP address range $I$}, in a given local network, is the union of the following IP address ranges:
\begin{enumerate}
    \item For an IPv4 address:
    \begin{itemize}
        \item The loopback space (127.0.0.0/8) defined in Section 3.2.1.3 of RFC 1122 \cite{rfc1122}.
        \item The link-local space 169.254.0.0/16 defined in RFC 3927 \cite{rfc3927}.
        \item The address space defined in Section 3 of RFC 1918 \cite{rfc1918}.
    \end{itemize}
    \item For an IPv6 address:
    \begin{itemize}
        \item The Local Address prefix (fc00::/7) defined in Section 3 of RFC 4193 \cite{rfc4193}.
        \item The link-local prefix (fe80::/10) defined in Section 2.5.6 of RFC 4291 \cite{rfc4291}.
    \end{itemize}
    \item The local network subnet address space, which might not be part of the previous items.
    \item The address spaces in an XML configuration file prepared by the network admin (see Subsection \ref{sec:implementation}).
\end{enumerate}
\end{mydef}

\begin{mydef}\label{external_ip_address_def}
The \textbf{external IP address space, $E$}, with respect to a local network, is all the IP addresses which are not in $I$.
\end{mydef}
\ignore{
\begin{mydef}\label{external_request}
A \textbf{cross-network HTTP request}, is an HTTP request which was initiated by a resource with an external IP address, and accesses a resource with an internal IP address.
\end{mydef}
}

Figure \ref{fig:INP_decision_process}, outlines the process of cross-network request handling by a web browser that follows the INP security policy. Whenever the browser detects a cross-network request, it first checks whether a cross-network request with the same parameters (external source IP, internal destination IP, HTTP headers and method), has been already approved by a corresponding preflight exchange. If this cross-network request has not yet been approved, then the browser sends a preflight request to the server at the internal destination IP address of the request. According to the preflight response, the browser either blocks or permits the cross-network request.

\begin{center}
\begin{figure}[h]
   \centering
   \begin{tabular}{@{}c@{\hspace{.5cm}}c@{}}
       \includegraphics[page=14,width=\linewidth]{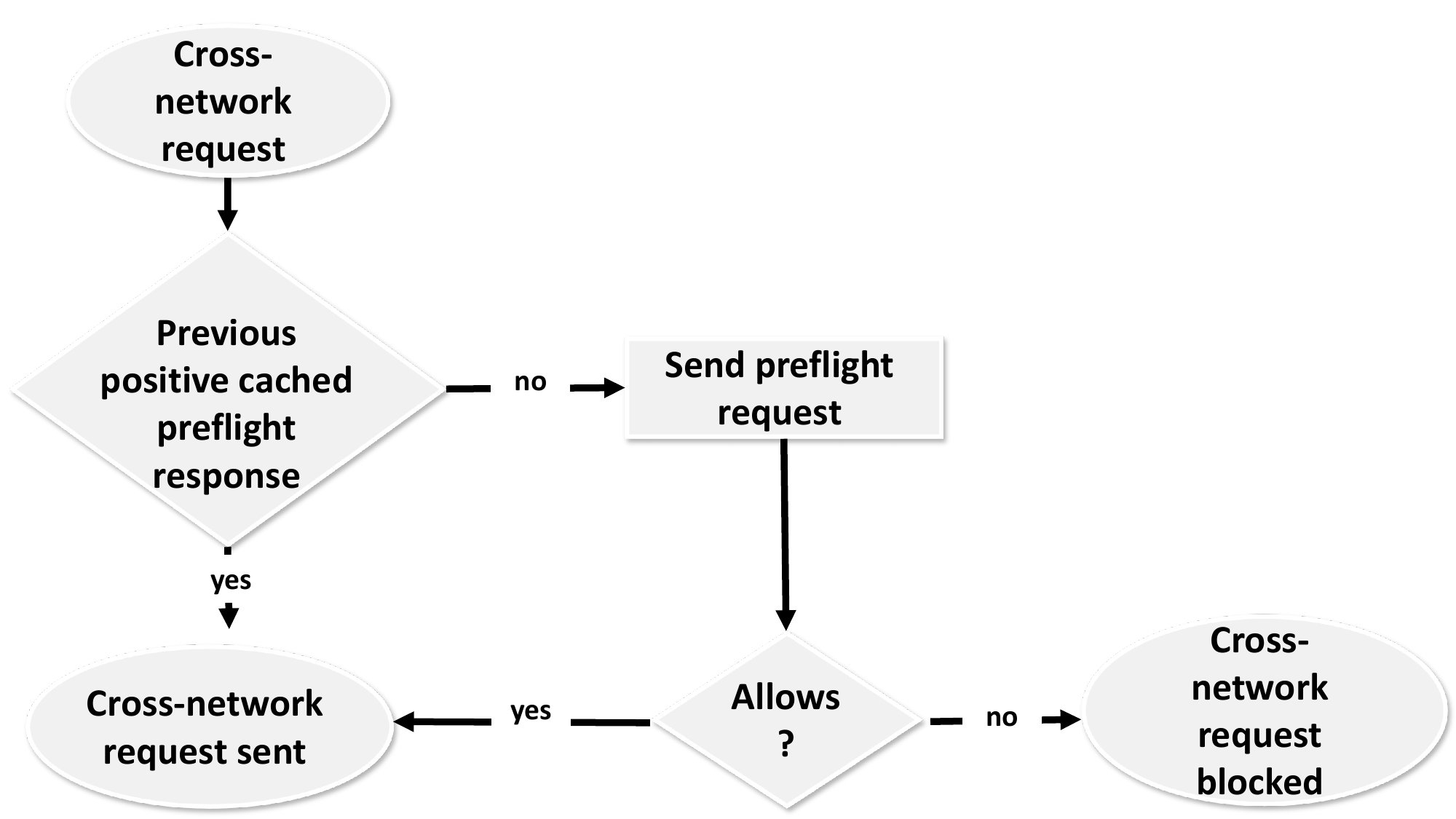}
   \end{tabular}
 \caption{INP process}
 \label{fig:INP_decision_process}
\end{figure}
\end{center}

\ignore{As SOP tracks a request's origin, INP tracks its initiator's IP address. The IP address space classification will be described later in the implementation details (Subsection \ref{sec:implementation}). }

A browser that implements INP keeps track of the source IP address of each page and resource it receives, and the destination IP address of each access it makes to detect any cross-network access event (some implementation details are given in Subsection \ref{sec:implementation}).

When a cross-network access event is detected and no corresponding approval has been issued to it yet, a preflight exchange with the destination internal server is initiated. Table \ref{table:INP_headers} provides the HTTP headers used in the preflight request and response. The request headers provide the internal resource detailed information about the external entity that has initiated this cross-network access (web origin and IP address) and the method and customized headers of the cross-network request. The internal server responds with a preflight response setting the headers to indicate whether it permits the cross-network request or not. An additional header in the response is X-INP-TTL, which gives the number of seconds the response to the preflight request can be cached and used without sending a new preflight request.

\begin{table*}[h!]
\centering
\begin{center}
\begin{tabular}{ |p{7cm}|p{7cm}| }
 \hline
 \multicolumn{2}{|c|}{\textbf{INP Headers}} \\
 \hline
 \textbf{INP preflight request headers} & \textbf{INP preflight response headers}\\
 \hline
 X-INP-Source-Origin & X-INP-Allow-Origin\\
 X-INP-Source-IP & X-INP-Allow-IP\\
 X-INP-Request-Method & X-INP-Allow-Methods\\
 X-INP-Request-Headers & X-INP-Allow-Headers\\
   & X-INP-TTL\\
 \hline
\end{tabular}
\end{center}
\caption{INP HTTP headers introduction}
\label{table:INP_headers}
\end{table*}


\ignore{In this section we explain the details of our suggested policy, INP.
The basic idea of INP is similar to SOP. At a high level, while SOP applies restrictions between web origins, INP applies restrictions between external and internal network resources. In the same way preflight requests are used by SOP to treat complex requests (as explained in Subsection \ref{sec:sop}), INP treats cross-network requests.}

\subsection{Security Evaluation} \label{secEval}
INP is designed to block cross-network accesses by default. Only accesses that obtain explicit permission from the internal destination are permitted. Notice that devices that do not implement INP preflight handling are protected by default, "no response'' to the preflight request is considered as a deny response.

Therefore, INP blocks by default attacks on devices behind a NAT that use the browser as a stepping stone, including all those we have presented in this paper, and many others.
In all the four attacks that were discussed in Section \ref{sec:attacks} an external page, evil.com, made a cross-network access to an internal device. INP detects the cross-network access in all these attacks and blocks it because it does not receive a positive response to the preflight request.

We demonstrate how INP prevents the previously shown attacks:
\begin{itemize}
    \item \textbf{Background attack - Acar et al.} - after DNS Rebinding occurs, the HTTP request, which was initiated by the script from evil.com with an external address, accesses an IoT device with an internal address. Therefore this request is a cross-network request. The browser would then send an INP preflight request to the device, which would probably not accept it. For that reason, the attacker's cross-network request would not be sent.
    \item \textbf{Attack I} - INP acts here just as the same as the case of Acar et al. attack. Instead of sending the INP preflight request to an IoT device, it would be sent to the home router.
    \item \textbf{Attack II} - the attack script from evil.com, triggers a request to an internal device. This is of course a cross-network request, which causes INP to send a preflight request to the device.
    \item \textbf{Attack III} - in this case, we show that INP handles request which were not only initiated by a dynamic script, but also from static HTML elements. INP would prevent the two steps which are required for the attack:
    \begin{itemize}
        \item Step of HTML form - which sends the login request to the home router which is classified as a cross-network request.
        \item Redirection - the attacker attempts to redirect the victim's browser from evil.com to the home router. However, this redirection is also a cross-network request, as its originated from evil.com with an external address and it is target is the home router whose address is an internal one.
    \end{itemize}
\end{itemize}

\subsection{Functional Evaluation}

INP is backward compatible, it only adds functionality in the cross-network event and otherwise the browser behaves the same as without INP.

One of the major advantages of INP is the fact that it is implemented in the web browser. This is critical as web browsers are frequently updated \cite{browsers_update} which is not the case with most devices such as routers supplied by ISPs and IoT devices. Thus all internal devices are protected by INP without any updates to the devices themselves.  

A major concern during designing INP was to avoid breaking any existing functionality in internal network resources, such as IoT devices. For example, INP allows by default an internal resource, to send HTTP requests through the user's browser to other internal resources. There are systems, such as smart homes, which use this internal access as part of their functionality. As will be described in our experiments (Section \ref{sec:experiments}), we checked if there is any access by design, from an external entity (e.g., a cloud server) to an internal network IoT device, through the web browser. We did not find any. Additionally, if a user wishes to surf directly to an internal network resource (such as an IoT web interface), it is recognized as internal to internal and is allowed by INP.

INP and SOP are naturally integrated together. In cases where a cross-network request is also a cross-origin complex one (this term was discussed in Subsection \ref{sec:sop}), then the browser sends a preflight request with both INP and SOP relevant headers. When the preflight response is received, both INP and SOP validations are required.

\subsection{Practical Implementation} \label{sec:implementation}
In order to validate the feasibility, security and functionality properties of the INP, we implemented a PoC (Proof of Concept) of INP in the Chromium web browser \cite{chromium_source}. Following are some of the practical challenges we overcame in the implementation of INP in Chromium.

\ignore{To detect cross-network requests the browser maintains for each page it is processing/displaying an initiator IP address. For each request that is made, the {\em destination} IP address of the request is determined. If the initiator address is external and the destination address of the request is internal then a cross-network event handling is invoked.}
To detect cross-network requests, for each request that is made, the browser maintains the IP of the request initiator (the IP of the website which created that request) and determines the destination IP address of the request. If the initiator address is external and the destination address of the request is internal, then a cross-network event handling is invoked.

In the PoC in Chromium, we were lucky to find the initial work of Mike West \cite{cors_rfc1918} (see Section \ref{sec:related}) to track whether an HTTP request is cross-network or not. However, we had to deal with several key differences: (i) our definition of internal is dynamic and more general, (ii) we had to add the preflight request and response handling, with all the extra parameters that are passed along (such as the initiator IP address), and (iii) in many scenarios the implementation in \cite{cors_rfc1918} makes the decision before the request destination has been resolved and thus might miss a cross-network request.

\ignore{\textbf{Tracking the IP address of the request's initiator}: The IP address of the initiator of the page in which the browser is executing is simply taken from the socket through which the browser got the page code (HTML and scripts). Conceptually, the request's initiator IP address extraction in INP is similar to the web origin extraction in SOP, and follows similar steps.} 
\textbf{Tracking the IP address of the request initiator}: Every request which is made by the browser, has an initiator, the web page within which the request is made. The request's initiator's IP address is simply taken from the socket through which the browser reached to the initiator (e.g., the web page). Conceptually, the request's initiator IP address extraction in INP is similar to the web origin extraction in SOP, and follows similar steps.  

\ignore{In the proof of concept in Chromium, we were lucky to find the initial work of Mike West \cite{cors_rfc1918} (see Section \ref{sec:related}) to track whether an HTTP request is cross-network or not. However, we had to deal with several key differences: (i) our definition of internal is dynamic and more general, (ii) we had to add the preflight request and response handling, with all the extra parameters that are passed along (such as the initiator IP address), and (iii) in many scenarios the implementation in \cite{cors_rfc1918} makes the decision before the request destination has been resolved and thus might miss a cross-network request.}    

\textbf{Extracting the request destination IP address}: A naive simple way would be to take the IP address after any required DNS resolution, just before the request is about to be made. However, due to the way Chromium is coded, the offered modifications by Mike West \cite{cors_rfc1918} check for cross-network request before some of the request destination domains are resolved. Thus it might miss a cross-network request because its destination IP address has not yet been resolved. Such cases still leave the door open for DNS rebinding attacks. To overcome this issue we added in our PoC a DNS resolution before each point where a cross-network detection is taking place and the destination is still given as a domain name and not as an IP address. Notice that this leaves a few microseconds (small number of machine instructions) between the time INP determined the destination IP address and the time the browser resolved the request IP address and has determined whether the request is cross-network. Thus theoretically a DNS rebinding attack could still take place with extremely small probability, if it is done in these few microseconds. Notice this is only a PoC implementation. In the real implementations of INP we suggest to make the cross-network detection after all the browser resolutions took place. 
\ignore{It is important not to resolve any HTTP request object which is created by the code of Chromium, as it would have dramatically affect the performances of the solution.}

\ignore{\textbf{Handling cross-network requests whose destination is not an IP address}: We added code which handles cross-network requests that target DNS names. In the current Chromium implementation, requests are determined cross-network before the DNS resolution takes place by the browser. Therefore, a cross-network request whose destination host is a DNS name of an internal network device, would not be considered cross-network, as the browser still does not know the IP address of the request's destination. Thus attacks as DNS Rebinding would not be prevented by the current modifications offered in Chromium \cite{cors_rfc1918}. In order to solve this case, we researched the process which HTTP requests pass through, from being initiated until their host is being resolved and the request is actually sent. Then, in specific critical points, we added a preceding DNS resolution of the host, in case it is not an IP address. It was extremely important not to resolve any HTTP request object which is created by the code of Chromium, as it would have dramatically affect the performances of the solution.
}

\textbf{Address space classification}: To dynamically acquire the internal network address space the browser can simply use the API of the machine it is running on (Windows, Linux, etc.) to get the network interfaces, IP addresses and subnet masks (both IPv4 and IPv6). For compatibility with larger organizational networks, which contain more than one address space, INP offers the network administrators the option to manually configure the organizational internal network address spaces. This can easily be implemented with an XML configuration file, which spreads in the network through management protocols such as GPO \cite{gpo}. An example of an XML configuration file is shown in Figure \ref{fig:xml_networks} in Appendix \ref{sec:appendix2}.

More technical details about our PoC implementation can be found in Appendix \ref{sec:appendix2}.

\section{ Experiments }
\label{sec:experiments}

In this section, we review the experiments we have done during our research. These include the setup of our three presented attacks, along with testing of our INP proof of concept.

\subsection{ Attack I - Attacking home routers with DNS Rebinding \& UPnP }

In order to simulate the attack execution, we used an Amazon Linux server on EC2 to host the attacker's malicious website (referred to earlier as evil.com). For the DNS Rebinding simulation, we set the victim's DNS server to be the Amazon server on EC2. The routers we chose to test the attack on are the most popular among the largest ISPs in our region, as verified with 183 students and colleagues. We tested the whole cycle with each vendor, from surfing to the attacker's website to attempting to send the UPnP commands to the router.

Most of the routers which enable UPnP in the LAN are vulnerable to our attack. Routers that are not vulnerable usually include a unique identifier in the UPnP URL (such us UUID). As the attacker does not have direct access to the router's LAN, the identifier cannot be acquired; thus, the UPnP server cannot be accessed. 

By sending an SSDP (Simple Service Discovery Protocol) M-SEARCH \cite{ssdp_msearch} request to a router, its UPnP server (if available) would respond with the server URL. This provides the attack code enough information, whether the router is vulnerable or not.

Table \ref{table:vuln_attack1}, presents the router distribution and the results of the experiments. More specific details about the vendors and the exact exploitation (such as the UPnP URL) can be viewed in Appendix \ref{sec:appendix1}. As can be seen in Table \ref{table:vuln_attack1}, 75\% of the tested routers are vulnerable to the attack.
\begin{table*}[h!]
\centering
\begin{center}
\resizebox{\textwidth}{!}{\begin{tabular}{ |c|c|c|c|c| }
 \hline
 \multicolumn{4}{|c|}{\textbf{Tested Routers}} \\
 \hline
 \textbf{Router vendor} & \textbf{Popularity} & \textbf{Is vulnerable to Attack II?} & \textbf{Is vulnerable to Attack IV?}\\
 \hline
 Sagemcom Fast & 22\% & \cellcolor{green!25}V & \cellcolor{green!25}V\\
 VTech & 14\% & \cellcolor{green!25}V & \cellcolor{red!25}X\\
 TP-Link & 23\% & \cellcolor{green!25}V & \cellcolor{red!25}X\\
 ADB Broadband Fast & 16\% & \cellcolor{green!25}V & \cellcolor{red!25}X\\
 D-Link & 19\% & \cellcolor{red!25}X & \cellcolor{green!25}V\\
 \hline
\end{tabular}}
\end{center}
\caption{Attacks II \& IV experiments results}
\label{table:vuln_attack1}
\end{table*}

\subsection{ Attack II - Attacking internal devices without DNS Rebinding }

The setup for this attack is quite simple, as we only use an Amazon Linux server to host our malicious website. We examined two IoT devices at home and created a proper attack script to control them.

\textbf{ Yamaha Network AV Receiver} 
The Yamaha RX-V683 Network AV Receiver exposes a web management interface on port 80. An attacker can send specific commands to the URL: "/YamahaRemoteControl/ctrl" to take control of the device. The commands are sent in HTTP POST requests without any authentication or customized headers needed. The body of the request is an XML (although no Content-Type HTTP header is necessary). Beside the web management interface which, listens on port 80, we discovered that the Yamaha RX-V683 Network AV Receiver exposes an interface which listens on port 49154.
Therefore, after scanning the internal network for a device which listens on this unique port, we sent the relevant HTTP request to control the device. 

\textbf{Sony Smart TV}
An attacker can send commands (HTTP POST request) without authentication directly to the HTTP web server of a Sony Smart TV with the URL: "/sony/IRCC?". Just as in the case of Yamaha, the command itself is an XML which appears in the body of the HTTP request. Sony supports many different commands, such as Power Off, Volume Control and Display Control, etc.

\subsection{ Attack III - Attacking home routers through static HTML elements }

Beside the website we host at Amazon, which acts as evil.com, in this attack, we also used our smartphone to set up a mobile hotspot. We tested the same routers as in Attack I. A simple experiment of creating a hotspot (with an HTML script tag as its name) and surfing to the router's AP List page can indicate whether the router is vulnerable to XSS. 
We used the network name \begin{verbatim}<script>alert("HACKED")</script>\end{verbatim} which popped up a message box in the victim's browser (see Figure \ref{fig:dlink_hacked}) when surfing to the AP List page. We tested the full attack flow, rather than this vulnerability alone. 41\% of the tested routers were vulnerable to the attack. The results are presented in Table \ref{table:vuln_attack1}.
\begin{figure}[h]
   \centering
   \begin{tabular}{@{}c@{\hspace{.5cm}}c@{}}
       \includegraphics[page=11,width=.45\textwidth]{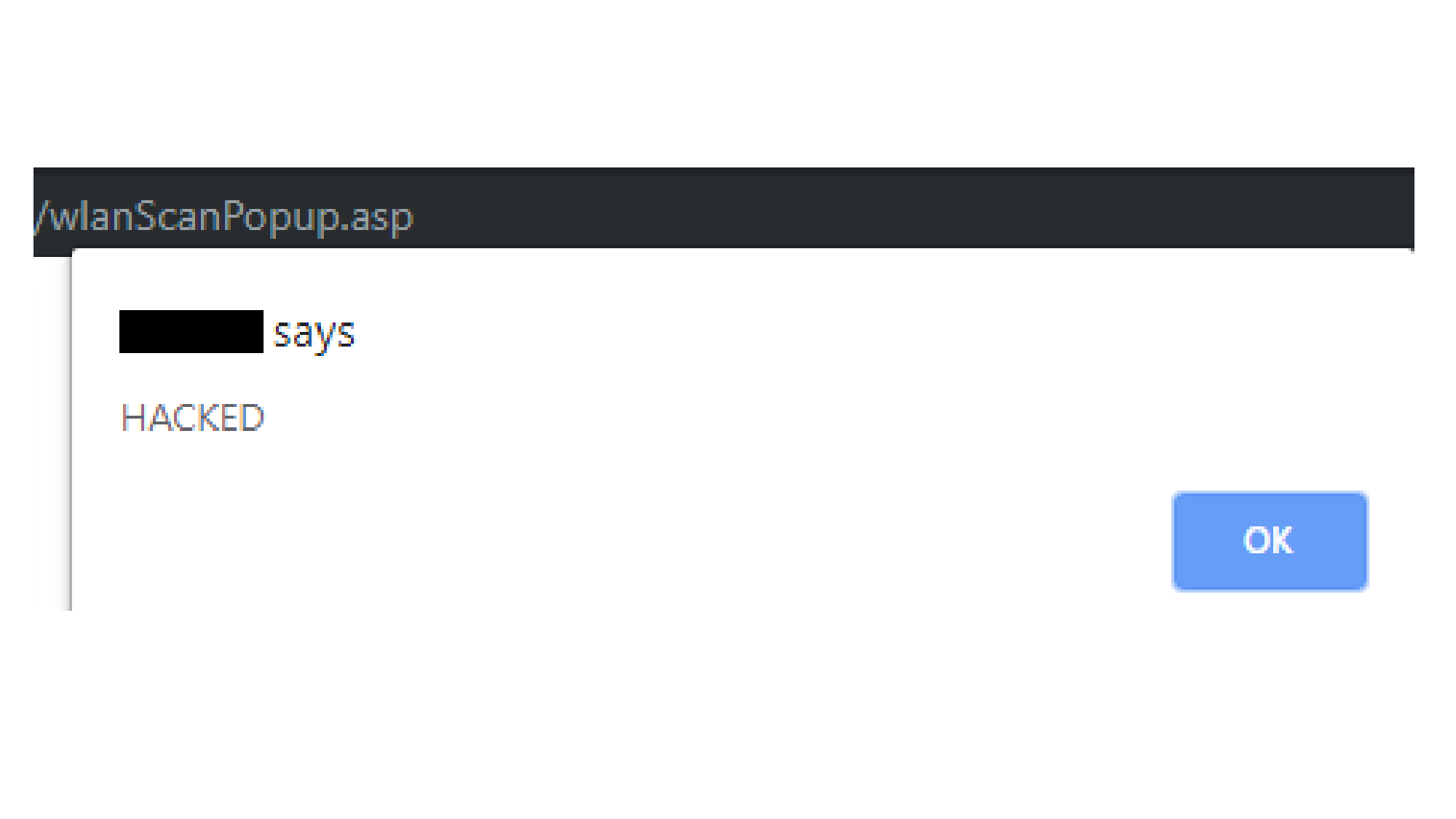}
   \end{tabular}
 \caption{Indication for a vulnerable Sagemcom router}
 \label{fig:dlink_hacked}
\end{figure}

\subsection{ Vulnerability disclosure }
We reported the vulnerabilities discovered throughout our research to the respective router (D-Link, TP-Link, VTech, Sagemcom, ADB) and IoT (Yamaha, Sony) vendors. Some of their security response teams are in contact with us, working together to fix the vulnerabilities.

\subsection{ Testing INP }
\subsubsection{Security}
As described in the INP implementation (Subsection \ref{sec:implementation}), to provide a proof of concept, we developed our version of Chromium, which applies INP. We tested the effectiveness of INP against our attacks in our compiled version of Chromium on a Windows 10 machine.
Our INP implementation successfully prevented all of our attacks by sending a preflight request to the internal network target instead of the attacker's malicious cross-site request.
\subsubsection{Performance}
We also examined the performance difference between the INP version of Chromium and the original one.
As discussed earlier, the implementation manifests itself mostly as parsing and extraction of the HTTP headers and string comparison in the address space classification. The comparison is executed for any request, and the header parsing is only for INP preflight requests.
In order to handle requests whose destination host is not an IP address, but a domain name, we added a preceding DNS resolution step, as explained in Subsection \ref{sec:implementation}.
In the beginning, our DNS resolution occurred at every processing by Chromium of any HTTP request. This dramatically lowered the browser's performance. Websites which were usually loaded by Chromium after a few seconds, were loaded by our browser version after a minute.
Then, we made a significant improvement by removing most of the times where the DNS resolution took place, keeping only the required for correctly determining cross-network requests (technical details can be viewed in Appendix \ref{sec:appendix2}).
We had practically no noticeable overhead when accessing a web application in our tests.
\subsubsection{Functionality}
While designing INP, we supposed that there is no legitimate access by design from an external entity (e.g., a cloud server) to an internal network IoT device, through the web browser. As part of our experiments, we used Fiddler \cite{fiddler}, a web debugging proxy, to extract HTTP (and HTTPS) traffic from workstations in home networks. In the extracted sessions, we looked for references for IP addresses in the workstation's LAN, or for DNS hosts which are not external web servers. Fortunately, we did not find any of those.
\section{ Availability }
\label{sec:availability}

We release samples of the attacks presented in this paper \cite{GIT_ATTACKS} along with our PoC implementation of INP with the Chromium web browser \cite{GIT_INP}, in a dedicated Github repository.

As part of the responsible vulnerabilities disclosure process, we do not release specific vendors' details online, but present the HTML pages, which evil.com would send the victim's browser in the  demonstrated attacks.

\section{Conclusion}
\label{sec:future}
In this paper, we have spotted the key vulnerability in most of the attacks behind a NAT, which use the web browser as a stepping stone. We realized that no matter which attacks are used, the browser would eventually be instructed by an external entity to access an internal network resource. Many studies have been conducted over the years \cite{johns2013eradicating,jackson2009protecting,barth2008robust}, most of them trying to prevent specific issues, while the root problem still exists. We have shown three attacks, each of which presents a different perspective of the problem. We then presented Internal Network Policy (INP), which deals with and solves the root cause and provides protection to users and organizations from being attacked and penetrated.

{\footnotesize 
\bibliographystyle{IEEEtran}
\bibliography{sample}

}

\begin{appendices}
\section{ Attacks }
\label{sec:appendix1}


\subsection{ Attack I - Attacking home routers with DNS Rebinding \& UPnP }
\subsubsection{Attacking randomized port UPnP servers}
There are routers, such as ADB Broadband or Asus, whose UPnP port is randomized at every boot of the device. According to standard convention \cite{iana}, the dynamic port range is 49152-65535, which means 16384 possible ports. However, there are routers such as ADB whose random UPnP port can be even lower then the specified above range. According to our experiments, Fetch API \cite{fetch} can be used to scan ten  thousands ports efficiently, with the ”no-cors” mode. Due to that, the attacker may add an additional step, scanning the router and then targeting the relevant discovered UPnP port. In our experiments, we succeed achieving a random UPnP port, in about 8 minutes.
\subsubsection{Vulnerable devices UPnP URLs}
See Table \ref{table:vuln_attack1_details} for vulnerable home routers and the corresponding URLs that invoke the required UPnP command:

\begin{table*}[h!]
\centering
\begin{center}
\begin{tabular}{ |c|c|c| }
 \hline
 \multicolumn{3}{|c|}{\textbf{Vulnerable Routers}} \\
 \hline
 \textbf{Router vendor} & \textbf{UPnP server TCP port} & \textbf{UPnP Control API URL} \\
 \hline
 Sagemcom Fast & 80 & /WANIPConnection\\
 V-Tech & 49600 & /upnp/control/WANIPConn1\\
 TP-Link & 1900 & /ctl/IPConn or /upnp/control/WANIPConn1\\
 ADB Broadband & Random & /ctl/WANPPPConnectionUPNP\\
 \hline
\end{tabular}
\end{center}
\caption{Routers vulnerable to attack I}
\label{table:vuln_attack1_details}
\end{table*}

The specific versions that we checked are:
\begin{itemize}
    \item \textbf{Sagemcom:} Fast 3184, Fast 3284
    \item \textbf{V-Tech:} NB403, IAD604, IAD605D
    \item \textbf{TP-Link:} VR400, VR600, W8970, W9970, Archer C7, Archer C9
    \item \textbf{ADB Broadband:} VV2220, VV5823
\end{itemize}

\subsection{ Attack II - Attacking internal devices without DNS Rebinding }
\subsubsection{Internal IoT devices identification}
In order to succeed, the attacker should identify an IoT device unique fingerprint, which could be detected while scanning the internal network, using the victim's web browser. As mentioned, this scanning should not utilize DNS Rebinding. We give some conceptual fingerprints and their detection methods.
\begin{itemize}
\item Unique open port on the device, can be discovered with port scanning \cite{git_beef}.
\item DNS name of the device, can be discovered with host scanning \cite{git_beef}.
\item Scanning specific unique URL which the device exposes \cite{img_scan}.
\end{itemize}

\subsection{ Attack III - Attacking home routers without using scripts }

\subsubsection{Scanning using static HTML tags}
Web servers, including those in home routers, tend to expose static web resources, such as images and scripts, without any authentication. In most of our attacks scenarios, this fact lets an external attacker identify what is the gateway vendor (and even version and firmware), only by trying to access common known static resources. As has been presented in a previous project \cite{img_scan}, an attacker may prepare a fingerprints database in advance, to successfully identify home routers.

\subsubsection{Wireless AP List}
There are routers whose web management interface exposes features such as AP List. This component provides the user wtih all the access points that are reachable to the router. The panel contains a basic table, with a few columns including the WiFi network name. The network name parameter is obtained directly from the access point itself.

\subsubsection{Routers Session Management}
There are mainly two different session management methods used in routers, and probably in most web servers: cookies or IP based sessions.
By using cookies, after the user successfully enter valid credentials, the router's web server sets a session cookie, which the browser uses during the session. In IP based session management, the user's session (based on his IP address) is saved in the router. In order to gain access to web pages, such as the AP List panel, all a user needs to do is send the relevant HTTP request, within the current active session.

\section{ INP }
\label{sec:appendix2}

\subsection{INP Chromium implementation}

\subsubsection{Current implementation}
In the source code of the Chromium project \cite{chromium_source}, the object that is responsible for managing HTTP requests and tracking their status, is ResourceRequest. This object includes many fields of a request, such as URL, origin, etc. The field that provides SOP with the information required to decide if a request should be restricted or not, is request\_initiator. For every HTTP request, the request\_initiator tells what is the origin, of the resource which initiated the current request.

The ResourceRequest has two classes. The first one takes place in the rendering process, which is responsible for processing data received from the network traffic, and display web pages to the user. The second one is in the browser process, and is responsible for sending and receiving data to/from the network, including capabilities such as sending HTTP requests and performing DNS resolution.

The method in the code which determines HTTP requests as cross-network (in the current offered modifications given in \cite{cors_rfc1918} they are call external), is SetExternalRequestStateFromRequestorAddressSpace. This method is implemented in the ResourceRequest class of the rendering process. However, in case where an HTTP request is about to be sent by the browser, and its destination is not an IP address but a domain name, the DNS resolution occurs only in the browser process, as explained earlier. 
This means that if the method SetExternalRequestStateFromRequestorAddressSpace, in the rendering process, handles an HTTP request whose destination is a domain name, the check would fail, and the request would not be considered as cross-network. Thus attacks as DNS Rebinding would not be prevented by the current modifications given in  \cite{cors_rfc1918}.

\subsubsection{INP implementation}

We briefly mention here the most critical changes we have made in the source code of Chromium, to properly integrate INP.

\textbf{Address space classification}: 
SetExternalRequestStateFromRequestorAddressSpace eventually checks whether the destination of an HTTP request, is in the browser's local network or not. The method in the code that performs this check, is called IsPubliclyRoutable. Because the original method only compared an IP address to static address spaces, we overrode it, and added comparison of the IP to the local network of the browser (using the running machine utilities such as NETSH in Windows) and to an XML file for organizations which can be configured by network admins (see Figure \ref{fig:xml_networks} for an example).
\begin{figure}[h]
   \centering
   \begin{tabular}{@{}c@{\hspace{.5cm}}c@{}}
       \includegraphics[page=5,width=.45\textwidth]{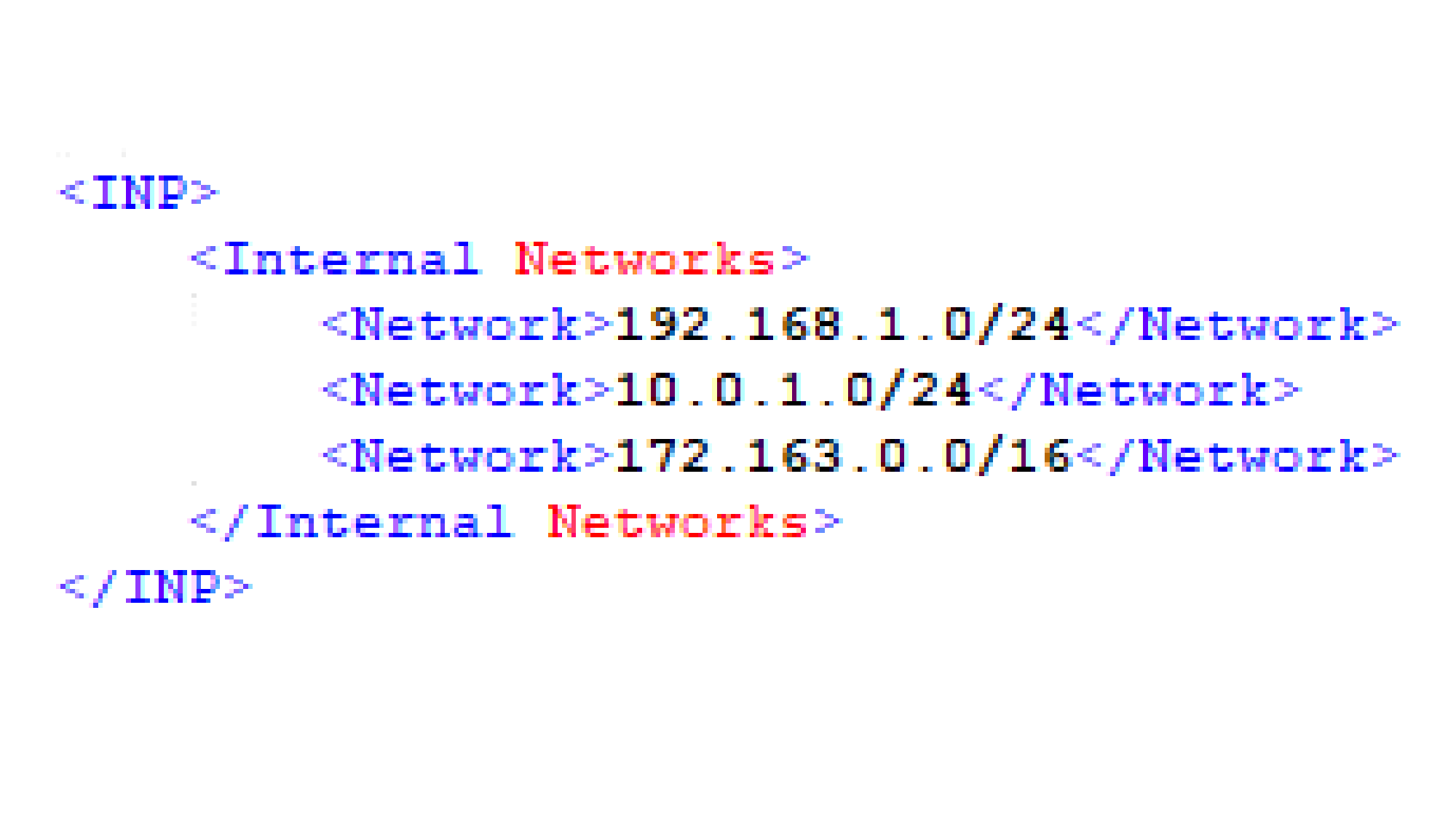}
   \end{tabular}
 \caption{INP Configuration example}
 \label{fig:xml_networks}
\end{figure}

\textbf{Handling cross-network requests whose destination is not an IP address}: 
In order to prevent attacks which target hosts whose address given as a domain name, instead of IP addresses, we had to address the challenge mentioned earlier. Because the DNS resolution occurs only in the browser process, we have to add an additional resolution step, which is executed by the rendering process. Two challenges had to be overcome in this part of the implementation:
\begin{enumerate}
    \item \textbf{Process permissions} - The rendering process is sandboxed and do not have any permission to execute network actions such as DNS resolution. Therefore, we implemented a helper process, that executed a DNS resolution on demand, and added a specific rule in the code which allowed sandboxed process, to run the helper process.
    \item \textbf{Performance} - In the beginning, we executed the DNS resolution at every call of SetExternalRequestStateFromRequestorAddressSpace. When we tested our compiled browser, we realized that the performance dropped as pages whose loading time was originally one second, where loaded only after a minute in our version. After a debugging session, we understood that SetExternalRequestStateFromRequestorAddressSpace is called many times per every HTTP request. Therefore, we moved the DNS resolution step from this method, to the code of CorsURLLoader, which checks for every HTTP request if SetExternalRequestStateFromRequestorAddressSpace marked the request as cross-network, and eventually handles the preflight requests. This change has improved the performance, and, in our tests we had no noticeable overhead when accessing a web application.
\end{enumerate}
\subsubsection{Future work}
In our current proof of concept, we only handle cross-network requests which target IPv4 addresses.
Additionally, we do not support the preflight caching mechanism, so every cross-network request will cause the browser to send a preflight request, even if a previous positive response was received recently.

\end{appendices}
ֿֿ
\end{document}